\documentclass[a4paper,11pt]{article}
\usepackage{adjustbox}
\usepackage{rotating} 
\usepackage{nameref} 
\usepackage{anyfontsize} 
\usepackage{float} 
\usepackage{graphicx} 
\usepackage{indentfirst} 
\usepackage{tabu} 
\usepackage{epstopdf} 
\usepackage{url} 
\usepackage{hyperref} 
\usepackage[table,xcdraw]{xcolor}
\usepackage{tikz} 
\usepackage[toc,page]{appendix}
\usepackage{hyperref} 
\usepackage{booktabs} 
\usepackage{threeparttable} 

\title{One Leak Will Sink A Ship: WebRTC IP Address Leaks} 
\author{Nasser Mohammed Al-Fannah \\ Information Security Group \\ Royal Holloway, University of London \\ \href{mailto:nasser@alfannah.com}{nasser@alfannah.com}}
\date{}

\begin{document}
	\maketitle
	\begin{abstract}
		The introduction of the WebRTC API to modern browsers has brought about a new threat to user privacy.   This API causes a range of client IP addresses to become available to a visited website via JavaScript even if a VPN is in use.   This a potentially serious problem for users utilizing VPN services for anonymity.  In order to better understand the magnitude of this issue, we tested widely used browsers and VPN services to discover which client IP addresses can be revealed and in what circumstances.  In most cases, at least one of the client addresses is leaked.  The number and type of leaked IP addresses are affected by the choices of browser and VPN service, meaning that privacy-sensitive users should choose their browser and their VPN provider with care.  We conclude by proposing countermeasures which can be used to help mitigate this issue. 
	\end{abstract}
	
	\section{Introduction}
	Ideally, when a user connects to the Internet via a Virtual Private Network (VPN), the IP addresses (e.g.\ the public IP address) of the client device are hidden from visited websites.  If a user is using a VPN for anonymity reasons, then revealing one, or more, of their IP addresses to a visited website or any browser add-on that can execute JavaScript on the client's browser is likely to negate the purpose of VPN use.  Revealing client IP address(es) could enable tracking and/or identification of the client.  Moreover, by using geolocation lookup, a client's public IP address could disclose its country and city \cite{geolocation}.  
	
	The introduction of WebRTC to modern web browsers has created a new and simple method for a visited website to discover one or more of the client IP addresses.  WebRTC is a set of APIs and communications protocols that provide browsers and mobile applications with Real-Time Communications (RTC) capabilities\footnote{\url{https://webrtc.org} [accessed on 14/05/2017]}. Apparently, identifying one or more of the client IP addresses via a feature of WebRTC was first reported and demonstrated by Roesler\footnote{The report and the demonstration script can be found at \url{https://diafygi.github.io/webrtc-ips} [accessed on 24/05/2017]} in 2015.  In this paper we refer to the WebRTC-based disclosure of a client IP address to a visited website when using a VPN as a \textit{WebRTC Leak}.
	
	The method due to Roesler can be used to reveal a number of client IP addresses via JavaScript code executed on a WebRTC-supporting browser.  Private (or internal) IP address(es) (i.e.\ addresses only valid in a local subnetwork) can be extracted from the exchange of a Session Description Protocol (SDP) object, which is necessary to establish a P2P (peer-to-peer) connection \cite{scanner}, while public (or external) IP address(es) (i.e.\ globally unique addresses) can be retrieved by successfully pinging a STUN server.  A STUN server (i.e.\ a Session Traversal of User Datagram Protocol Through Network Address Translators (NATs) server) allows a NAT client to set up interactive communications, such as a phone call, to a VoIP provider hosted outside the local network \cite{scanner}.
	
	In this paper, we describe experiments performed to examine five types of client IP address that could be revealed via the WebRTC functionality.  We also examined to what degree the choice of browser, VPN service and VPN client-side configuration affects the number and type of leaked addresses.  A related investigation has been described by Perta et al.\ \cite{glass}, who observed the role of the VPN service in IP address leaks.  However, they focused only on IPv6 address leaks without looking at other types of IP address or the role of the browser in the leaks.  Moreover, the address leaks considered are apparently not WebRTC-related.  This is the first paper to examine all the types of IP address that could leak, as well the first to consider the role of the browser in these leaks.
	
	It is important to note that WebRTC leaks could affect client privacy even if a VPN is not in use.  This is because the client private IP address could be leaked, a piece of information which would not otherwise be available to a visited website even in the absence of a VPN.\ However, these addresses are not necessarily very privacy-sensitive, since clients are typically assigned private IPv4 addresses in the 192.168.0.x range \cite{augment}.
	
	The remainder of the paper is structured as follows.  In section 2 we discuss the types of IP address that could potentially be leaked via WebRTC.  We review prior work related to WebRTC leaks in section 3.  The research methodology employed as well as details of the experiments performed are discussed in sections 4 and 5.  In section 6, we report on and analyse the results of these experiments.  Before concluding in section 8, we discuss WebRTC leak countermeasures in section 7.
	
	\section{IP Addresses At Risk}
	In the experiments (see Section 5) we found that WebRTC functionality can be exploited to reveal one or more of five types of client IP address, as listed below.  Note that the public IPv4 address of a client is not in the list, as in the experiments we performed we were never able to learn such an address using WebRTC.  
	
	\begin{itemize}
		\item \textit{Public IPv6 address}: this is the IPv6 address of the platform and is typically assigned by the ISP of the client.
		\item \textit{Public Temporary IPv6 address}: this address is assigned by the network to which the client platform is attached.
		\item \textit{Unique local address (ULA) assigned by LAN}: this IPv6 address is assigned by the network to which the client platform is attached, and is the approximate IPv6 counterpart of the \textit{Private IPv4 address assigned by LAN} \cite{ula}.
		\item \textit{Private IP address assigned by the VPN server}: this private (IPv4 or IPv6, depending on the VPN configuration) address is assigned by the VPN server.
		\item \textit{Private IPv4 address assigned by LAN}: this address is assigned by the network to which the client platform is attached.
	\end{itemize}
	
	The disclosure of an IPv6 address is more privacy-damaging than that of the private IPv4 address.  Moreover, the Public IPv6 address remained the same throughout more than two months of testing while the temporary IPv6 address changed with every connection instance.  However, the persistence of an IP address depends on the client and network configuration, but there is no doubt that the public IPv6 address is more persistent than a temporary IPv6 address (hence the name temporary).  
	
	More generally, the degree to which the disclosure of a particular type of IP address degrades user privacy depends on its uniqueness and persistence.  For example, a private IPv4 address (4 bytes) is typically in the 192.168.0.x range, and is thus far less privacy-sensitive than a public IPv6 address (16 bytes).  Moreover, a leak of the IP addresses of clients that are assigned static (i.e.\ fixed) IP addresses will be more privacy-compromising than if these addresses are dynamically assigned (i.e.\ they change regularly).
	\section{Previous Work}
	WebRTC leaks have been discussed in several previous studies \cite{augment, million, scanner, p2p, tracing, glass}.   Jakobsson \cite{p2p} explores WebRTC leaks in the greatest depth, but focuses only on public IP address leaks.  
	
	Alaca et al.\ \cite{augment} observed that WebRTC features could enable a visited website to learn the IP addresses assigned to all the network interfaces of a client platform, including the private IP addresses assigned by a VPN.  They deemed this possibility to be a medium-level threat, which seems a reasonable evaluation given they only observed the possibility of private IP address leaks.  However, they state in their evaluation that the WebRTC leak issue requires further study.  
	
	Englehardt et al.\ \cite{million} consider the WebRTC threat in some depth, but like many other authors they only examine private IP address leaks.  Liu et al.\ \cite{tracing} only examine leaking of private IP addresses.  They also claim that the WebRTC issue is only applicable to Chrome and Firefox, but it is not clear what browsers and which versions they tested (the results we obtained, described in Section \ref{exper}, contradict this claim).
	
	Hosoi et al.\ \cite{scanner} point out that public IP addresses could be amongst those leaked.   As previously mentioned, Perta et al.\ \cite{glass} explore IPv6 address leaks in detail.  They report on the results of IP address leak tests of 14 VPN services; however, they do not describe the role of WebRTC in these leaks.
	
	A recent draft RFC, entitled \textit{WebRTC IP Address Handling Requirements} \cite{ietf}, details browser mechanisms that can potentially prevent WebRTC-related IP address leaks.
	
	In summary, a number of authors have examined the WebRTC issue, but none have made a comprehensive survey of the issue; typically they have either only examined some of the possible IP addresses that can be leaked, or not considered the roles of both the browser and the VPN service in affecting the magnitude of the leaks.  In the remainder of this paper we describe the results of the first comprehensive study of the WebRTC leak issue, including examining the roles of the browser, VPN service and VPN configuration in affecting the nature and volume of IP addresses leaked.  This enables us to make recommendations to end users on how they might optimise their behaviour to minimise their loss of privacy.  We have also provided a website which enables users to test the privacy properties of their own current configurations.
	\section{Experimental Methodology}
	
	We used a modified version of Roesler's publicly available JavaScript to perform the experiments.  The modification incorporates some of the features provided by BrowserLeaks.com that enable the script to work with Edge, which Roesler's original script does not support.  Preliminary tests revealed that the number and type of leaked addresses are affected by the choices for both the web browser and the VPN service.  We therefore tested five different widely used VPN services running on eight different browser-OS combinations, namely four browsers each running on Windows and macOS.  Since we had no access to any publicly available information regarding VPN services that are most widely used, we informally selected five of the top search results in Google.  The VPN services we chose to examine are: Hide My A**!\ (HMA!), ZenMate, ExpressVPN, VyprVPN and TorGuard.  Full details of the OSs and VPNs used in the experiments can be found in Appendix A.
	
	For browsers, we chose to examine the five most widely used desktop programs according to netmarketshare.com\footnote{\url{https://www.netmarketshare.com/browser-market-share.aspx} [accessed on 14/05/2017]}, namely \textbf{Chrome, Firefox, Edge, Safari} and \textbf{Opera}. Although Internet Explorer is the second most widely used browser, we excluded it from the study because it does not support WebRTC and so is not affected by the leaks discussed in this paper.  Moreover, it has been replaced by Edge as the default browser in Windows.  Since Chrome, Firefox and Opera are available on Windows and macOS, we tested these three browsers on both OSs. 
	
	Most of the tested VPN programs provide means for the users to modify some VPN configurations to, for example, switch from one VPN protocol (e.g.\ L2TP) to another (e.g.\ PPTP).  We found that in some cases this also affects which IP addresses are leaked.  This is likely to be due to the VPN server configuration rather than the protocol itself.  Nevertheless, this fact is important to recognize and so we indicate in our results summary below the VPN programs that exhibited such differences (see Appendix A for details of tested VPN configurations).
	\section{Details of Experiments} \label{exper}
\begin{figure}[!t]
		\centering
		\includegraphics[width=5in]{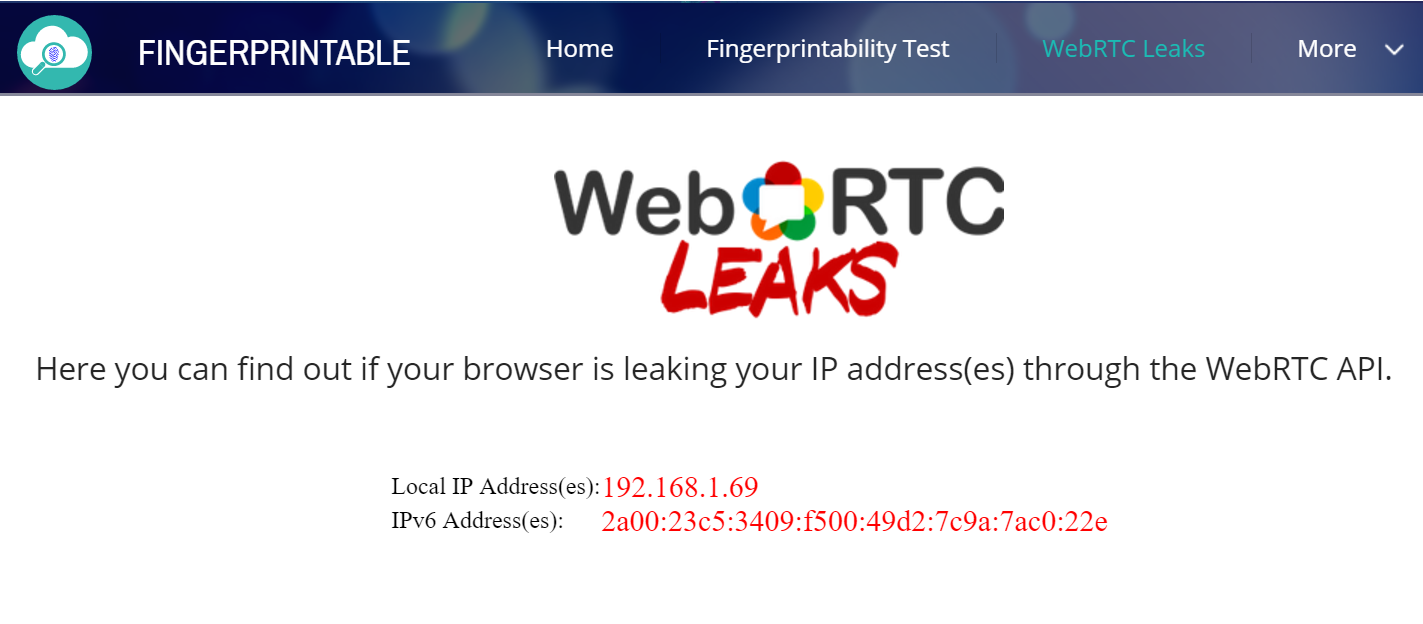}
		\caption{WebRTC leak detector}
		\label{screenshot}
	\end{figure}

	To perform the experiments, a website (\url{https://fingerprintable.org/webrtcleaks}) was specially established.  The web page contains JavaScript that, when executed in a client browser, fetches all the IP addresses it can retrieve using WebRTC; the leaked IP addresses (if any) are then displayed on the page (see Figure \ref{screenshot}).  When using it for the tests, the visiting device used either the Windows \textit{ipconfig} command at the command prompt, or \textit{ifconfig} in macOS terminal to identify the types of address displayed on the page.
	
	We deployed all five of the chosen VPN programs with each of the eight selected browser-OS combinations, giving a total of 40 test cases.  In each case, we caused the client to visit the test page on the specially established website, and documented the IP address(es) displayed.  For each of the 40 (VPN, OS, browser) combinations, we visited the test page using all the protocols/configurations supported by the VPN to detect any differences in leaked IP addresses, i.e.\ for each of the 40 test cases we made between one and five tests (for full details see Appendix B), giving a total of 116 tests.  For example, for each browser-OS combination we visited the test page using VyprVPN a number of times, one time using L2TP/IPsec once using PPTP, and so on.

	The tested VPN services provide access to VPN servers in a range of countries.  However, in a series of informal tests we found no difference in the set of leaked IP addresses when connecting to VPN servers for the same service in different countries.
	\section{Results and Analysis}
	Tables 1 and 2 summarize the experimental results for Windows and macOS, respectively.  Listed in the tables are the types of IP addresses leaked in each test environment.  The VPN protocols are also given in the cases where the choice of protocol made a difference to the set of leaked IP addresses.  It is worth noting that tests on macOS while deploying VPN did not reveal client private IPv4 address, public IPv6 address or ULA.

	\subsection{VPNs}
	The choice of VPN service had a significant effect on the number and type of IP addresses leaked.  In some cases, using one VPN protocol (e.g.\ L2TP/IPsec) in a VPN program leaked a different number of addresses than another protocol in the application.  We observed no differences in address leakage when switching between the TCP and UDP network protocols, in VPN services that enabled such options for users.  However, when testing different VPN programs, variations in the sets of leaked IP addresses were observed even when the same protocol was in use.  We therefore concluded that these differences can be attributed to how the VPN service is configured to handle the connection when using particular protocols.   
	
	As can be seen from the tables, TorGuard proved to be the least privacy-compromising VPN service.  In all test cases, it revealed none of the client's public IP addresses.  At the other extreme, VyprVPN and ExpressVPN did not prevent any of the WebRTC leaks.
	
	\begin{center}
		\begin{table}[H]
			\caption{Results of experiments on Windows }\label{tab1}
			\fontsize{8}{8}\selectfont
			{
				\begin{tabular}{|l|p{4.8em}|p{4.8em}|p{4.8em}|p{4.8em}|}
					\hline
					\rowcolor[HTML]{ EFEFEF} 
					VPN / Browser & Chrome & Firefox & Edge & Opera
					\\ \hline
					Without VPN & pvt.\ IPv4 \newline IPv6 & pvt.\ IPv4 & pvt.\ IPv4 \newline IPv6 \newline ULA & pvt.\ IPv4 \newline IPv6
					\\ \hline
					\toprule
					\bottomrule
					
					HMA! (all protocols) & VPN IPv4 & VPN IPv4 & VPN IPv4 \newline pvt.\ IPv4 \newline IPv6 \newline ULA & VPN IPv4
					\\ \hline
					\toprule
					\bottomrule
					
					ZenMate & VPN IPv4 \newline temp.\ IPv6 & VPN IPv4&VPN IPv4 \newline pvt.\ IPv4 \newline IPv6 \newline ULA & VPN IPv4 \newline temp.\ IPv6
					\\ \hline
					\toprule
					\bottomrule
					
					ExpressVPN (all protocols) & VPN IPv4 \newline temp.\ IPv6 & VPN IPv4 & VPN IPv4 \newline pvt.\ IPv4 \newline IPv6 \newline ULA & VPN IPv4; temp.\ IPv6
					\\ \hline
					\toprule
					\bottomrule
					VyprVPN (Chameleon \& OpenVPN) & VPN IPv4 \newline temp.\ IPv6 & VPN IPv4 & pvt.\ IPv4 \newline IPv6 \newline ULA & VPN IPv4 \newline temp.\ IPv6
					\\ \hline
					VyprVPN (L2TP/IPsec \& PPTP) & temp.\ IPv6 & no leak & pvt.\ IPv4 \newline IPv6 \newline ULA & temp.\ IPv6
					\\ \hline
					\toprule
					\bottomrule
					TorGuard (OpenVPN) & VPN IPv4 & VPN IPv4 & pvt.\ IPv4 \newline VPN IPv4 & VPN IPv4
					\\ \hline
					TorGuard (OpenConnect) & VPN IPv4 & VPN IPv4 & no leak & VPN IPv4
					\\ \hline
					
				\end{tabular}
			}
			\begin{tablenotes}\footnotesize
				\item \textbf{IPv6} = public IPv6 address; \textbf{temp.\ IPv6}=public temporary IPv6 address; \textbf{ULA} = unique local address; \textbf{VPN IPv4} = private IP address assigned by VPN server; \textbf{pvt.\ IPv4} = private IPv4 address assigned by LAN.
			\end{tablenotes}
			
		\end{table}
	\end{center}
	\subsection{Browsers}
	Safari revealed no IP addresses, and this is because its WebRTC-support remains under development.  This is clear from the Safari browser engine (namely webkit) specifications status page\footnote{\url{ https://webkit.org/status/} [accessed on 14/05/2017]}.   By contrast, Edge revealed four of the five IP addresses discussed in this paper (only the temporary IPv6 address was not leaked).  Edge was the only browser to reveal the public IPv6 address(es) and ULA(s).  This makes it the most privacy-damaging browser.  This might be because at the present Edge is the only browser that supports the next generation WebRTC API, named \textit{ORTC} (Object Real-Time Communications)\footnote{\url{https://ortc.org/faq} [accessed 14/05/2017]}
	
	Opera and Chrome were identical in terms of the number and type of leaked addresses.   This is likely because both are based on Google's open-source browser project, \textit{Chromium}\footnote{ \url{http://www.chromium.org/blink/developer-faq} [accessed on 14/05/2017]}.  In all the individual tests that resulted in IP address leakage, they both revealed the temporary IPv6 address and either the local private IP address or the VPN-assigned private IP address.  Somewhat different behaviour was exhibited by Firefox, which in most cases revealed either the local private IP address or the VPN-assigned private IP address; in some cases it did not reveal any addresses. 
	
	Firefox was the least privacy-damaging of the Windows-based browsers and Edge the most.  In macOS, Safari revealed no IP addresses and so it is the least privacy-damaging.  Chrome and Opera were the most privacy-damaging macOS browsers.

	\begin{center}
		\begin{table}[H]
			\caption{Results of experiments on macOS }\label{tab2}
			\fontsize{8}{8}\selectfont
			{
				
				\begin{tabular}{|l|p{4.8em}|p{4.8em}|p{4.8em}|p{4.8em}|}
					\hline
					\rowcolor[HTML]{ EFEFEF} 
					VPN / Browser & Chrome & Firefox & Safari & Opera
					\\ \hline
					
					Without VPN & pvt.\ IPv4 \newline IPv6 & pvt.\ IPv4 & no leak & pvt.\ IPv4 \newline IPv6
					\\ \hline
					\toprule
					\bottomrule
					
					HMA! (PPTP) & VPN IPv4 & VPN IPv4 & no leak & VPN IPv4
					\\ \hline
					HMA! (OpenVPN) & VPN IPv4 \newline temp.\ IPv6 & VPN IPv4 & no leak & VPN IPv4 \newline temp.\ IPv6
					\\ \hline
					
					\toprule
					\bottomrule
					
					ZenMate & VPN IPv4 \newline temp.\ IPv6 & VPN IPv4 & no leak & VPN IPv4 \newline temp.\ IPv6
					\\ \hline
					\toprule
					\bottomrule
					
					ExpressVPN (OpenVPN) & VPN IPv4 \newline temp.\ IPv6 & VPN IPv4 & no leak & VPN IPv4 \newline temp.\ IPv6
					\\ \hline
					ExpressVPN (L2TP/IPsec) & VPN IPv4 & VPN IPv4 & no leak & VPN IPv4
					\\ \hline
					\toprule
					\bottomrule
					VyprVPN (Chameleon \& OpenVPN) & VPN IPv4 & VPN IPv4 & no leak & VPN IPv4
					\\ \hline
					VyprVPN (L2TP/IPsec) & no leak & no leak & no leak & no leak
					\\ \hline
					\toprule
					\bottomrule
					TorGuard (all protocols) & VPN IPv4 & VPN IPv4 & no leak & VPN IPv4
					\\ \hline
					
				\end{tabular}
			}
			\begin{tablenotes}\footnotesize
				\item \textbf{temp.\ IPv6}=public temporary IPv6 address; \textbf{VPN IPv4} = private IP address assigned by VPN server; \textbf{pvt.\ IPv4} = private IPv4 address assigned by LAN.
			\end{tablenotes}
			
		\end{table}
	\end{center}
	\section{Countermeasures}
	The main lesson from the experiments described in this paper is that users concerned about IP address leaks should select their browser and VPN service with care, perhaps using the \url{https://fingerprintable.org/webcrtleaks} site to check the properties of the chosen combination.  Over and above this, users interested in maintaining their privacy by preventing WebRTC leaks can perform one or more of the countermeasures discussed below.  In this context, the first countermeasure has previously been discussed by Hosoi et al.\ \cite{scanner} and the second countermeasure by Perta et al.\ \cite{glass}.  
	
	It is worth noting that disabling JavaScript would prevent WebRTC leaks but would also disable many features and functionality of modern websites.  Most users are likely to find this an unacceptably high cost for the privacy enhancement they would receive, in the same way that whilst disabling cookies has significant privacy benefits, the usability impact is too great to make it a widely used protection measure.
	\begin{itemize}
		\item \textbf{Disable WebRTC} Disabling WebRTC in a browser would prevent all of the leaks discussed in this paper.  Typically, this countermeasure can be implemented via the browser user settings. However, such an approach may not be acceptable to all users, since it will disable all the functionality provided by WebRTC.
		\item \textbf{Disable IPv6} Of course disabling IPv6 addressing in a client computer means there are no IPv6 addresses (both regular and temporary) or ULA to be leaked.  Private IPv4 address(es) can still be leaked but they are much less privacy-compromising than the other addresses. 
		\item \textbf{Anonymizing add-ons}  Browser add-ons are available that block leakage of client addresses.  For example, Chrome has an official add-on called \textit{WebRTC Network Limiter}\footnote{\url{https://chrome.google.com/webstore/detail/webrtc-network-limiter/npeicpdbkakmehahjeeohfdhnlpdklia}} that prevents WebRTC leaks.  However, as stated in the add-on menu, using the add-on could negatively affect WebRTC features.
		\item \textbf{Browser choice} As demonstrated in this paper the choice of browser makes a significant difference to the type and number of leaked IP addresses.  A user can check their browser of choice by visiting the test page to discover if IP addresses are leaked when using a VPN.
		\item \textbf{VPN choice} As demonstrated in this paper, WebRTC leaks also depend on the VPN in use and its configuration.  A user can test their VPN of choice for WebRTC leaks to help decide whether or not it meets their privacy needs.  Again, this can be performed by visiting the test web page and checking if any leaked IP addresses are displayed.
	\end{itemize}
	\section{ Summary and Conclusions}
	In the experiments performed in this study, Safari did not cause any client IP addresses to be leaked via WebRTC leaks.  Edge, on the other hand, proved to be the most privacy-damaging in this respect.  However, regardless of the user browser choice, we found that some VPN implementations prevent leakage of client public IP address(es).  Moreover, in some cases, selecting an appropriate client VPN configuration fully or partially prevented WebRTC leaks.
	
	The experiments we performed in this study explored an aspect of WebRTC leaks that has not been addressed in previous work, namely that the choice of browser and VPN service makes a significant difference to the extend of WebRTC leaks.   The results will help users decide on best practices to minimize the risk of WebRTC leaks.  We also hope it will encourage VPN and browser providers to work on mitigating the privacy-compromising properties of their implementations of the WebRTC API.

	\bibliographystyle{plain}
	\bibliography{vpn}

\begin{thebibliography}{1}

\bibitem{augment}
Furkan Alaca and Paul~C. van Oorschot.
\newblock Device fingerprinting for augmenting web authentication:
  classification and analysis of methods.
\newblock In Stephen Schwab, William~K. Robertson, and Davide Balzarotti,
  editors, {\em Proceedings of the 32nd Annual Conference on Computer Security
  Applications, {ACSAC} 2016, Los Angeles, CA, USA, December 5--9, 2016}, pages
  289--301. {ACM}, 2016.

\bibitem{million}
Steven Englehardt and Arvind Narayanan.
\newblock Online tracking: {A} 1-million-site measurement and analysis.
\newblock In {\em Proceedings of the 2016 {ACM} {SIGSAC} Conference on Computer
  and Communications Security, Vienna, Austria, October 24--28, 2016}, pages
  1388--1401. {ACM}, 2016.

\bibitem{ula}
R.~Hinden and B.~Haberman.
\newblock Unique local ipv6 unicast addresses.
\newblock RFC 4193, IETF, October 2005.

\bibitem{scanner}
Rio Hosoi, Takamichi Saito, Takayuki Ishikawa, Daichi Miyata, and Yongyan Chen.
\newblock A browser scanner: Collecting intranet information.
\newblock In {\em 19th International Conference on Network-Based Information
  Systems, NBiS 2016, Ostrava, Czech Republic, September 7--9, 2016}, pages
  140--145. {IEEE} Computer Society, 2016.

\bibitem{p2p}
Christer Jakobsson.
\newblock Peer-to-peer communication in web browsers using {WebRTC}: A detailed
  overview of {WebRTC} and what security and network concerns exists.
\newblock Bachelor's thesis, Department of Computing Science, Umeå University,
  2015.
\newblock
  \url{http://www8.cs.umu.se/education/examina/Rapporter/ChristerJakobsson.pdf}.

\bibitem{geolocation}
Robert Koch, Mario Golling, and Gabi~Dreo Rodosek.
\newblock Geolocation and verification of {IP}-addresses with specific focus on
  {IPv6}.
\newblock In {\em Cyberspace Safety and Security --- 5th International
  Symposium, {CSS} 2013, Zhangjiajie, China, November 13--15, 2013,
  Proceedings}, pages 151--170, 2013.

\bibitem{tracing}
Xiaofeng Liu, Qixu Liu, Xiaoxi Wang, and Zhaopeng Jia.
\newblock Fingerprinting web browser for tracing anonymous web attackers.
\newblock In {\em Proceedings of 1st IEEE International Conference on Data
  Science in Cyberspace (DSC), Changsha, China, June 13--16, 2016}, pages
  222--229. IEEE, 2016.

\bibitem{glass}
Vasile~Claudiu Perta, Marco~Valerio Barbera, Gareth Tyson, Hamed Haddadi, and
  Alessandro Mei.
\newblock A glance through the {VPN} looking glass: Ipv6 leakage and {DNS}
  hijacking in commercial {VPN} clients.
\newblock {\em PoPETs}, 2015(1):77--91, 2015.

\bibitem{ietf}
Guo wei Shieh and Justin Uberti.
\newblock {WebRTC IP Address Handling Requirements}.
\newblock Internet-Draft draft-ietf-rtcweb-ip-handling-03, Internet Engineering
  Task Force, January 2017.
\newblock Work in Progress.
  \url{https://datatracker.ietf.org/doc/html/draft-ietf-rtcweb-ip-handling-03}.

\end{thebibliography}

	\begin{appendices}
		\section{Program Versions} 
		\subsection{Operating Systems}
		The following OSs were used in the experiments.
		\begin{itemize}
			\item Windows 10.0.14393 (Build 14393)
			\item macOS ‎10.12.4‎ (16E195)
		\end{itemize}
		
		\subsection{VPN Programs}
		The VPN program versions and URLs that were tested are listed below.
		\begin{center}
			\begin{table}[H]
				\caption{VPN Program Versions}\label{vpnver}
				\fontsize{9}{9}\selectfont
				{
					\begin{tabular}{|l|l|l|l|}
						\hline
						\rowcolor[HTML]{ EFEFEF} 
						VPN / Specs & Windows & MacOS & URL
						\\ \hline
						HMA! & 3.4.6.1 & 2.2.7.0 & \url{https://hidemyass.com}
						\\ \hline
						ZenMate & 3.4.7.17 & 1.5.4 & \url{https://zenmate.com}
						\\ \hline
						ExpressVPN & 6.0.9 & 6.3.3 & \url{https://expressvpn.com}
						\\ \hline
						VyprVPN & 2.9.6.7227 & 2.14.0.5485 & \url{https://goldenfrog.com/vyprvpn}
						\\ \hline
						TorGuard & 0.3.69 & 0.3.69 & \url{https://torguard.net}
						\\ \hline

					\end{tabular}
				}		
			\end{table}
		\end{center}
		
		\subsection{Browsers}
		Listed below are the names and versions of browsers used in the experiments.  The version numbers are the same on both Windows and MacOS.
		\begin{itemize}
			\item Chrome 58.0.3029.110
			\item Firefox 53.0.2
			\item Edge 38.14393.1066.0
			\item Opera 45.0.2552.635
			\item Safari 10.1 (12603.1.30.0.34)
			
		\end{itemize}
		\section{Tested VPN Configurations}
		
		\begin{center}
			\begin{table}[H]
				\caption{Tested VPN program configurations}\label{config}
				\fontsize{8}{8}\selectfont
				{
					\begin{tabular}{|l|p{7em}|p{3.5em}|p{7em}|p{4.8em}| p{8.4em}|}
						\hline
						\rowcolor[HTML]{ EFEFEF} 
						OS / VPN & HMA! & ZenMate & ExpressVPN & VyprVPN & TorGuard
						\\ \hline
						Windows & 
						OpenVPN UDP \newline OpenVPN TCP & 
						N/A & 
						OpenVPN UDP \newline OpenVPN TCP \newline L2TP/IPsec \newline PPTP \newline SSTP &
						Chameleon* \newline OpenVPN \newline L2TP/IPsec \newline PPTP  &
						OpenVPN UDP \newline OpenVPN TCP \newline OpenConnect UDP\newline OpenConnect TCP 
						\\ \hline
						macOS & 
						OpenVPN \newline PPTP & 
						N/A & 
						OpenVPN UDP \newline OpenVPN TCP \newline L2TP/IPsec &
						Chameleon* \newline OpenVPN \newline L2TP/IPsec  &
						OpenVPN UDP \newline OpenVPN TCP \newline OpenConnect UDP\newline OpenConnect TCP 
						\\ \hline
						
					\end{tabular}
				}		
				\begin{tablenotes}\footnotesize
					\item *VyprVPN proprietary protocol 
				\end{tablenotes}	
			\end{table}
		\end{center}
	\end{appendices}
	
\end{document}